# Upstream Laser-based Longitudinal Enhancement of Relativistic Photoelectrons

Hao Zhang[1,2*], Randy Lemons[2], Jack Hirschman[2,3], Nicole Neveu[2], Nicolas Sudar[2], River Robles[2,3], Paris Franz[2,3], David Cesar[2], Zihan Zhu[2], Mathew Britton[2], Kurtis Borne[2], Zhen Zhang[2], Kirk A. Larsen[2], Benjamin Mencer[2], Justin Baker[1], Chad Pennington[1], Razib Obaid[2], Yuantao Ding[2], Ryan Coffee[2], Gabriel Just[2], Feng Zhou[2], Ji Qiang[5], James Cryan[2], Joseph Robinson[2], Agostino Marinelli[2], Sergio Carbajo[1,2,6,7]

[1]Department of Electrical & Computer Engineering, UCLA, Los Angeles, CA 90095, USA
[2]SLAC National Accelerator Laboratory, Stanford University, Menlo Park, California 94025, USA
[3]Department of Applied Physics, Stanford University, CA 94305, USA
[4]Department of Mathematics, UCLA, Los Angeles, CA, 90095, USA
[5]Lawrence Berkeley National Laboratory, Berkeley, CA 94720
[6]California NanoSystems Institute, 570 Westwood Plaza, Los Angeles, CA 90095, USA
[7]Physics and Astronomy Department, University of California, Los Angeles, CA 90095, USA
*haozh@ucla.edu; scarbajo@g.ucla.edu

**Abstract:** Controlling the longitudinal phase space of high-brightness relativistic electron beams is crucial for advancing a broad spectrum of charged-particle-based instrumentation and scientific frontiers. A generalized method for achieving this control involves manipulating the photoemission laser's temporal distribution at the picosecond level—a long-standing technical challenge. Recent developments in laser shaping have enabled the creation of high-power, picosecond-scale symmetrical and asymmetrical temporal profiles, capable of fine-tuning complex space-charge dynamics and external field effects in relativistic charged-particle beams. Here, we demonstrate that rather than deviations from theorized, idealized laser distributions, a controlled asymmetry can be harnessed to counteract accelerator-induced distortions. By implementing spatiotemporal shaping of the ultraviolet photocathode laser at the LCLS-II superconducting injector, we achieve deterministic control over the longitudinal phase space without downstream corrections. We find that this optical asymmetry induces a self-linearizing effect across both low (~40 pC) and high (80 pC) charge regimes, effectively suppressing nonlinear compression and energy chirp. Consequently, this approach is expected to preserve a low emittance comparable to that of ideal flattop or regular Gaussian profiles, while delivering superior current uniformity and shot-to-shot stability. These results establish spatiotemporal laser shaping as a compact, generalizable tool for directly optimizing beam brightness at the source.

**Keywords:** Longitudinal phase space dynamics, laser temporal shaping, space-charge effect, relativistic electron beams, photoinjector physics, X-ray free-electron lasers

## Main

The increasing demands on beam brightness, stability, and compression across diverse relativistic electron applications call for more deterministic, source-level control strategies.[1–3]. High-brightness electron beams are foundational to a wide range of applications, including X-ray free-electron lasers (XFELs)[4,5], ultrafast electron diffraction (UED)[6,7], inverse Compton scattering (ICS)[8,9], attosecond science[10–12], electron microscopy[13,14], high-energy physics[15,16], strong field QED[17], and advanced radiation sources spanning THz to gamma-ray regimes[18,19]. In all these platforms, the electron beam's initial phase space characteristics, especially its emittance and longitudinal structure, strongly influence the quality, efficiency, stability, and full electromagnetic structure of the emitted radiation[20–25]. As facilities move toward higher repetition rates and more compact configurations, upstream control becomes increasingly critical to minimize beam losses and eliminate the need for complex downstream manipulation[26].

A uniform and extended high-current region within the electron bunch is essential for stable and efficient radiation generation in beam-driven light sources[27–29]. The portion of the bunch where the instantaneous current remains near its peak, the effective current window, determines how much charge contributes coherently to the FEL gain process[30]. A flatter and broader current profile improves gain uniformity, enhances beam–wave coupling, and reduces timing sensitivity, all of which are critical for maintaining spectral coherence and pulse stability[31]. In seeded FELs, such profiles promote stronger overlap and phase locking between the seed field and the electron beam, improving spectral purity[32,33]. In time-resolved experiments, including attosecond streaking and ultrafast pump–probe studies, a smooth current profile ensures consistent temporal structures[11,34–36]. Flatter profiles support more efficient bunch compression and help suppress microbunching instabilities, particularly important for low-charge, high-repetition-rate operation[37]. Narrow or distorted current profiles limit usable charge, increase sensitivity to jitter, and amplify shot-to-shot variation, undermining the performance of next-generation light sources.

Beyond the width of this high-current region, the magnitude of the peak current itself also plays a central role. Stable high peak current is particularly important for high-charge FEL operation and other beam-driven radiation sources[38–41]. In FELs, higher peak current shortens the gain length, reduces the undulator distance to saturation, and enhances photon yield—benefits that are especially important for compact facilities and short-pulse operation[18,42–44]. In ICS sources for THz or γ-ray generation, higher current densities directly improve radiation brightness and enable tighter bandwidth control. Applications such as nonlinear X-ray spectroscopy[45], single-particle imaging[46,47], and ultrafast materials science[48,49] all rely on dense electron bunches capable of driving intense and temporally confined radiation fields. Achieving and maintaining high peak current with minimal distortion under realistic conditions, especially at high charge, is therefore a key challenge.

This requirement for phase-space linearity shifts the engineering focus back to the electron source. In state-of-the-art photoinjectors, the temporal structure of the electron beam is largely dictated by the intrinsic envelope of the ultraviolet (UV) drive laser and the emission physics at the photocathode[50–52]. Under high-brightness conditions, the rapid onset of space-charge forces couples with the native laser temporal profile to generate correlated curvature, energy–time chirp, and current nonuniformity in the emerging electron bunch[50,53–55]. These longitudinal features are subsequently amplified through compression and collective effects, and are only partially correctable downstream using harmonic RF linearization, wakefield structures, or beam-based tuning. While such methods have enabled modern XFEL performance, they increase operational complexity and do not provide direct control at the emission stage. This motivates an upstream approach in which the longitudinal phase-space structure is deterministically engineered at the source level[56].

Conventionally, the cylindrical flattop-temporal profile has been pursued as the canonical target for such source engineering, as its uniform current distribution theoretically minimizes the non-linear components of the space-charge field[3,27,28,52,57–59]. Generating such mathematically ideal distributions in practice is fundamentally constrained by both optical physics and material limitations[52,60]. A strictly rectangular temporal shape necessitates infinite spectral bandwidth; in real systems, frequency-domain truncation inevitably results in Gibbs oscillations and ringing[61,62]. Compounding these spectral limits is the formidable challenge of manipulating high-energy pulses in the UV regime. The picosecond timescale is too fast for direct electronic modulation, yet the high photon energy imposes severe damage threshold constraints on optical materials[3,60,63]. This fragility precludes the use of high-resolution adaptive shaping elements, such as liquid crystal spatial light modulators or complex diffractive optics, often employed in infrared systems[64]. Consequently, forced to rely on simpler, robust amplification schemes, the pulse profile becomes dominated by the thermodynamics of the laser medium. As energy is extracted, gain saturation (Frantz-Nodvik effect) naturally imparts a time-dependent skew to the pulse, resulting in a characteristic quasi-flattop profile with a trailing intensity decay[65]. Realistic photoemission is therefore driven not by ideal rectangles, but by asymmetric pulses that seed correlated distortions into the electron beam.

In this paper, rather than eliminating imperfections in the laser temporal profile, we investigate how a controlled asymmetry, specifically, a head-biased negative slope, can be used to counteract accelerator-induced distortions and improve current uniformity. We demonstrate that deterministic control over the longitudinal phase space of relativistic electron beams can be achieved purely through spatiotemporal shaping of the ultraviolet photocathode laser, without relying on downstream corrections or major accelerator retuning. Implemented at the superconducting LCLS-II injector, this technique yields beams with near-linear longitudinal energy–time correlations and low projected energy spread across both low (~40 pC) and high (80 pC) charge regimes. At low charge, the shaped beam maintains a stable and reproducible temporal profile, with a large fraction of charge concentrated in the high-current core. This helps reduce timing jitter and space-charge distortion during injector transport. At high charge, the same shaping suppresses nonlinear compression and limits energy–time curvature, resulting in lower energy chirp and better shot-to-shot consistency—all achieved without relying on a laser heater. For the first time, we demonstrate that spatiotemporal laser shaping provides a compact and generalizable tool for directly controlling the longitudinal phase space at the source, enabling improved beam quality and stability across a wide range of accelerator-based light sources.

## Results and Discussion

To demonstrate deterministic longitudinal beam control via spatiotemporal laser shaping, we implemented a programmable pulse-shaping system[52,66,67] at the LCLS-II photoinjector, as illustrated in Figure 1. The photoinjector drive laser shaping system begins with a programmable infrared (IR) modulation stage, where a spatial light modulator (SLM) applies spectral phase shaping to seed pulses before amplification, as shown in Figure 1 (inset ①). These pre-shaped pulses are then injected into a commercial IR amplifier. The amplified IR pulses are subsequently directed into a custom chirp-controlled UV conversion module, which we proposed in the previous work[27,52], that performs sum-frequency generation (SFG) followed by second-harmonic generation (SHG), as illustrated in Figure 1 (inset ②). The IR pulses are first split into two replicas, each independently chirped—one with a positive chirp and the other with a negative chirp. Additional amplitude shaping is applied to one arm via a wire mask at the Fourier plane. The oppositely chirped pulses are first spatially overlapped and mixed in a beta-barium borate (BBO) crystal to generate second-harmonic sum-frequency pulses at $2\omega$. These $2\omega$ pulses are then frequency-doubled in a subsequent BBO crystal to produce fourth-harmonic ($4\omega$) radiation with a controllable, flattened temporal intensity profile ideally suited for flat-current, high-brightness electron beam generation[27,52,67]. The iris acts as a spatial filter, selectively transmitting only the central region where sum-frequency generation occurs, while blocking the divergent SHG beams that do not contribute to the final UV profile. By adjusting both the chirp rates and spatial overlap conditions (e.g., crystal orientation), the spatiotemporal shape of the final UV output pulse can be precisely tuned to achieve the desired uniform, flattop profile with durations on the order of tens of picoseconds. These tens-of-picoseconds durations are chosen in the LCLS-II design to balance space-charge mitigation with RF synchronization, ensuring high-brightness electron bunch generation.

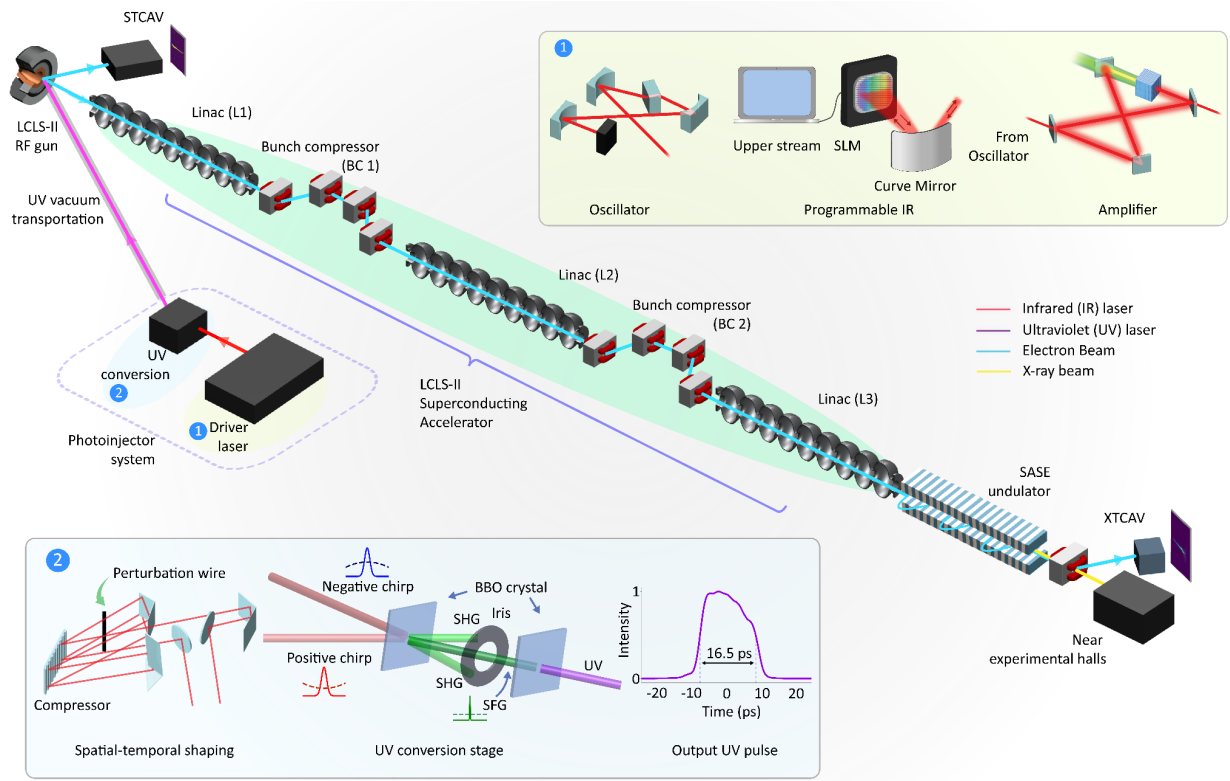

*Figure 1. Overview of the experimental beamline and laser shaping system at LCLS-II. (Main panel) Schematic of the experimental beamline (not to scale) from the programmable drive laser system to the undulator section. The shaped UV pulses generated at the photoinjector drive laser are used to photoemit electron bunches from the RF photocathode gun. The bunches are accelerated through the injector and linac (L1–L3), and characterized by multiple diagnostic stations: the STCAV downstream of the cathode, and the XTCAV downstream of the undulator, both of which provide longitudinal phase space and current measurements. (Inset ①) The drive laser shaping system begins with an SLM that pre-modulates the seed before injection into the Carbide. (Inset ②) The UV-shaping module operates on two IR pulse replicas with equal and opposite chirps, with additional amplitude shaping applied in one arm. The two replicas are combined in a BBO crystal to generate the desired non-collinear sum-frequency signal, while parasitic SHG is spatially filtered with an iris. This sum-frequency output is then frequency-doubled in a second BBO crystal to produce the flattop UV pulse, whose profile can be fine-tuned by adjusting chirp and crystal alignment.*

To characterize the longitudinal electron beam properties, we used a set of energy and time-resolved diagnostics positioned both upstream and downstream of the undulator. The S-band transverse deflecting cavity (STCAV), located upstream of the undulator, provides measurements of bunch length and phase space after the injector before further compression. Further downstream, the X-band transverse deflecting cavity (XTCAV), located after the undulator, enables measurement of the longitudinal phase space, X-ray temporal profile, and current profile at the end of the linac. These diagnostics allow us to track how laser shaping affects beam evolution from emission through acceleration to the undulator entrance and beyond. The shaped UV pulses were also temporally characterized via cross-correlation, confirming the generation of flattened temporal profiles with a full-width half-maximum (FWHM) near 17 ps.

**Longitudinal electron beam at the injector**

To evaluate the impact of spatiotemporal laser shaping on longitudinal beam dynamics, we performed a series of simulations and measurements at bunch charges of 40 pC and 80 pC. At lower charge, the beam is typically operated in a moderately compressed regime at LCLS-II, where current profile uniformity plays a key role in determining both brightness and temporal fidelity. To study the effect of shaping under stronger compression, we also carried out measurements at 80 pC. No laser heater was used during these experiments due to equipment constraints, which made the beam particularly sensitive to space-charge effects and microbunching instabilities[3,68,69]. While this increased sensitivity introduced challenges, it also allowed a direct evaluation of how initial laser shaping affects the evolution of longitudinal phase-space structure.

To investigate the sensitivity of beam quality to realistic temporal shaping imperfections, we performed a systematic study of quasi-flattop laser profiles with varying degrees of temporal asymmetry. In practice, ideal flattop pulses are difficult to realize due to gain depletion in the amplification chain and pulse reshaping during transport, often resulting in a current profile that is skewed or asymmetric at the photocathode. By introducing a controlled slope K in the temporal laser profile, we mimic this asymmetry and evaluate its effect on longitudinal beam dynamics.

Figure 2a shows representative laser profiles used in simulation: an ideal flattop and quasi-flattop pulses with K = –0.15 and –0.50. All profiles are normalized to the same FWHM and total charge. As the slope increases, the pulse becomes more asymmetric, and its overall temporal extent grows to conserve total area. Figures 2b and 2c present the resulting transverse emittance degradation and RMS bunch length broadening at 40 pC and 80 pC. The ideal flattop yields the lowest emittance, while quasi-flattop profiles with K ≈ –0.1 to –0.3 show minimal degradation. Beyond K = –0.4, emittance and bunch length increase significantly, primarily due to the extended pulse tail and enhanced space-charge effects.

Simulated longitudinal phase space distributions in Figs. 2d–f, h–j show broadly similar overall structure across different shaping conditions at both 40 pC and 80 pC. However, distinct differences emerge in the output current profiles (top projections). For the ideal flattop input, the current profile at the injector exit becomes skewed, with a stronger leading edge and a sloped trailing edge. Although the laser profile is temporally flat, the current asymmetry arises from longitudinal beam dynamics within the photoinjector. Electrons emitted at earlier times typically experience stronger accelerating fields and more effective longitudinal compression, due to a combination of RF phase curvature and space-charge effects. Electrons emitted later in the pulse may occupy less favorable RF phases or accumulate greater energy spread, leading to weaker compression and temporal broadening. As a result, even a symmetric input can evolve into a left-skewed current profile after acceleration. The quasi-flattop case with $K = -0.15$ partially compensates for this effect. The slight asymmetry in the laser profile offsets the non-uniform evolution in the injector, resulting in a nearly symmetric output current with steep rise and fall edges. For $K = -0.50$, the temporal profile becomes significantly broadened with lower peak current, consistent with the increased emittance observed in Fig. 2b and c.

We also performed experimental measurements under quasi-flattop shaping, as shown in Figs. 2g and 2k. The projected current and energy spread of the longitudinal phase space were measured using the upstream STCAV, under linac settings described in the Methods and Supplementary Information. The laser profile used in the experiment corresponds to a quasi-flattop shape with an asymmetry slope of approximately $K = -0.18$ (see Fig. 1).

The measured longitudinal phase space exhibits a narrow energy spread and a well-preserved, near-linear energy–time correlation, indicating good overall beam quality. The bunch maintains a high-current duration of ~8.5 ps (FWHM), or ~3.6 ps RMS, which supports stable beam loading. This profile structure is expected to help suppress shot-to-shot fluctuations and improve longitudinal consistency in the downstream linac and undulator. Interestingly, the current profile measured at the injector exit shows a similar left-skewed shape as observed in the simulation of the ideal flattop case (Figs. 2d and h), rather than the flatter output seen in the simulated quasi-flattop case (e.g., Fig. 2e and i). This confirms that even with a quasi-flattop laser input, the output current distribution is shaped significantly by the injector's longitudinal dynamics, as discussed earlier.

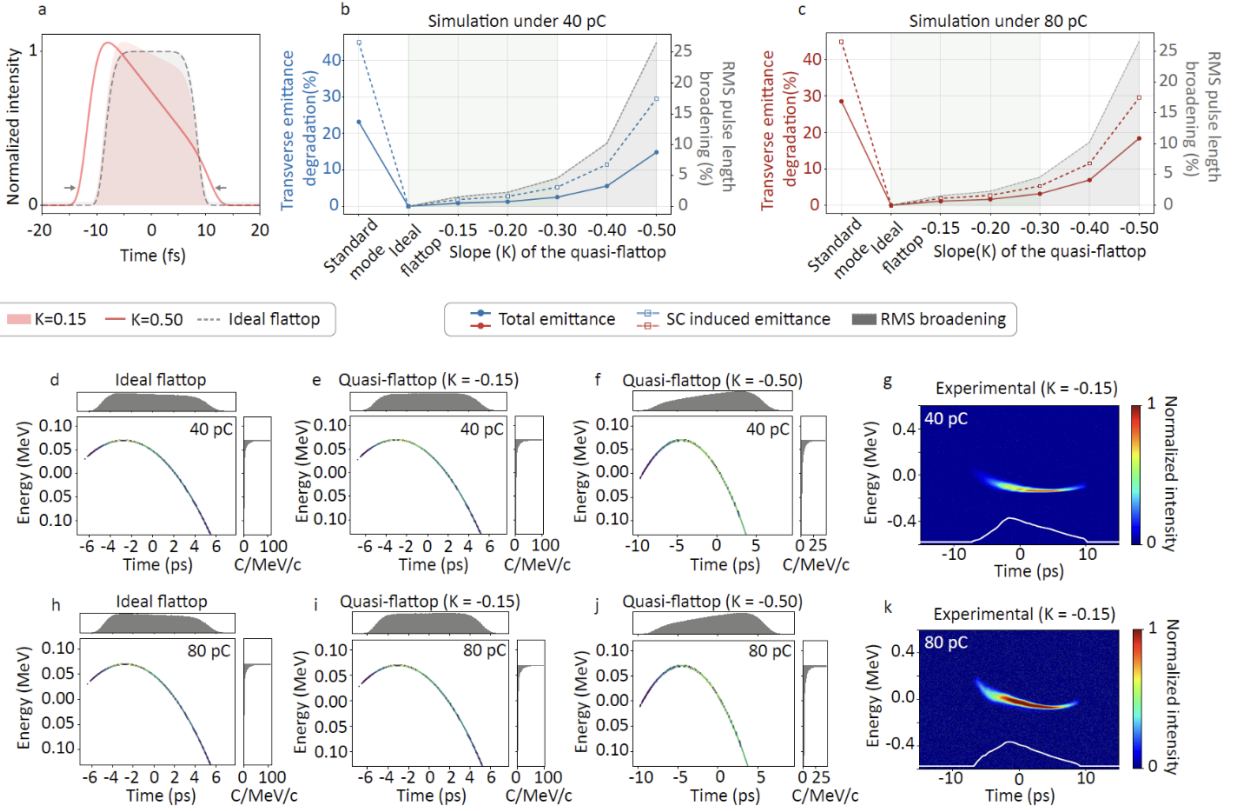

*Figure 2. Impact of quasi-flattop temporal shaping on longitudinal beam phase space and emittance. (a) Example of temporal laser profiles used in simulation, comparing the ideal flattop shape, quasi-flattop shapes with asymmetry slope K = –0.15 and K = -0.50. (b, c) Simulated emittance degradation (total and space charge–induced) and RMS bunch length broadening as a function of shaping slope K, at 40 pC (b) and 80 pC (c). Minimum emittance is observed around K ≈ –0.1 to –0.20. (d–f, h–j) Simulated longitudinal phase space distributions for different shaping conditions (ideal flattop, K = –0.15, and K = –0.50) at 40 pC (d–f) and 80 pC (h–j). (g, k) Measured longitudinal phase spaces and time-current under K ≈ –0.15 shaping at 40 pC and 80 pC.*

**Longitudinal beam characterization after the undulator**

To evaluate the downstream behavior of the quasi-flattop beam, we performed shot-resolved measurements at the end of the linac using the XTCAV. Figure 3 summarizes the results over 200 consecutive shots at a bunch charge of approximately 40 pC. Figure 3a shows the average current profile across all shots. Compared to the standard operating mode (see Supplementary Information), the quasi-flattop shaping results in a broader and flatter central region. The shaded gray area indicates the portion of the bunch where the instantaneous current exceeds 30% of the peak, which contains ~80% of the total charge. This higher charge concentration in the core may help improve beam loading uniformity and FEL gain stability.

Figure 3(b) presents a representative longitudinal phase space measurement from XTCAV. The energy centroid trace (white) is fitted with a global linear regression (orange dashed) and a local fit (red solid) within the high-current window, providing two complementary estimates of the energy chirp. In this example, the chirp appears small, and the energy spread remains tightly concentrated, indicating good longitudinal beam quality. Figure 3(c) shows the shot-to-shot variation in measured charge across the dataset, demonstrating consistent injector performance with only moderate fluctuations over time.

Figures 3d and e display the statistical distributions of extracted energy chirp values using global and local linear fits, respectively. Chirp values obtained under quasi-flattop shaping are reasonably small and stable across the sample.

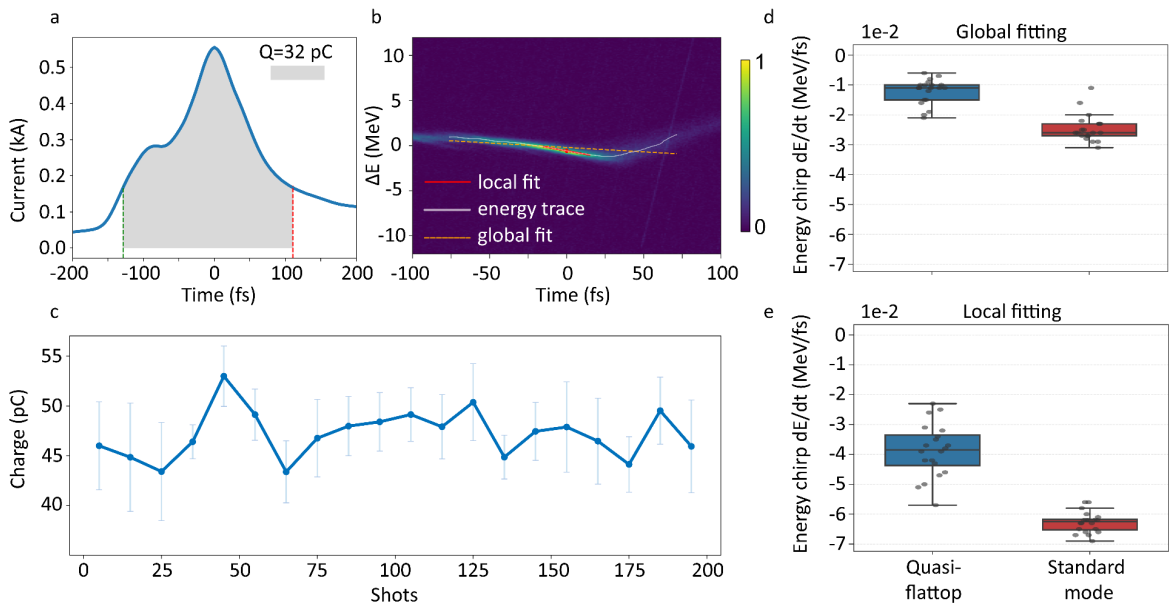

*Figure 3. Characterization of quasi-flattop beam and longitudinal chirp over 200 shots under low charge level. (a) Averaged current profile over 200 consecutive shots at 36 pC. The shaded gray region indicates the portion of the beam where the instantaneous current exceeds 30% of the peak, used as a window for chirp analysis. (b) Longitudinal phase space measurement of a representative shot after the undulator. Energy centroid trace (white) is fitted globally (orange dashed) and locally (red) to extract chirp. (c) Shot-to-shot variation of the measured charge over the normal operation. (d, e) Distributions of extracted energy chirp (dE/dt) from global (d) and local (e) linear fitting, comparing quasi-flattop and standard laser profiles. The comparison is made over the same number of shots and fitting window, using the laser profile as the benchmark variable.*

We extended the quasi-flattop shaping measurements to a higher bunch charge of 80 pC to evaluate the performance under more demanding space-charge conditions. Figure 4a shows the averaged temporal current profile over 60 consecutive shots. The high-current window contains approximately 80% of the total bunch charge on average, and up to 84% at 2 kA in representative single-shot examples (Figure 4b).

Figure 4c shows the measured longitudinal phase space of a representative shot at 80 pC. The energy centroid trace (white) is fitted globally and locally within the high-current region to extract the chirp, following the same method as in Figure 3. The observed chirp is small, and the energy spread is tightly localized. Figure 4d presents the shot-to-shot variation of the total charge across the measurement window, showing consistent injector operation with moderate fluctuations.

Figures 4(e) and 4(f) show the distributions of extracted energy chirp using global and local linear fits at 80 pC. Under quasi-flattop shaping, the chirp values are significantly lower than those in the standard mode, and the statistical spread is also narrower in both metrics. This indicates improved shot-to-shot consistency in addition to reduced energy–time distortion. In contrast, Figures 3(d) and 3(e) show that at 40 pC, quasi-flattop shaping still leads to lower chirp values compared to the standard mode, but the statistical spread is comparable in the global fit and slightly larger in the local fit.

This may be because at low charge, collective effects are weak, and the beam evolution is mainly governed by RF curvature and laser profile details. Without a laser heater, the beam is more vulnerable to microbunching and profile-induced shot-to-shot fluctuations. The steep edges of the quasi-flattop shape also make it more sensitive to temporal jitter or noise, contributing to the broader chirp spread. At high charge, space-charge and compression effects dominate. In this regime, the quasi-flattop's smoother current profile helps suppress nonlinear distortions and reduce energy–time curvature. The fact that this improvement appears even without a laser heater highlights the intrinsic stability advantage of quasi-flattop shaping under strong collective effects.

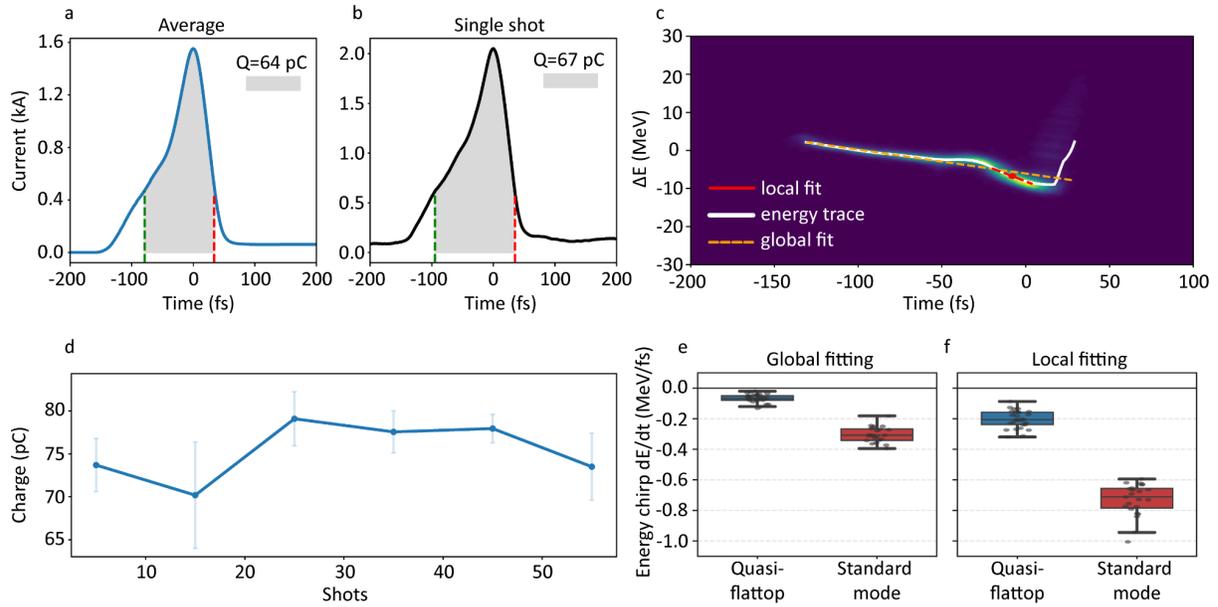

*Figure 4. Longitudinal phase space and energy chirp characterization under quasi-flattop shaping at high charge level. (a, b) Temporal current profiles for the quasi-flattop beam at higher charge. (a) The averaged profile over 60 shots, and (b) a representative single shot with 2kA at 80 pC. The shaded region corresponds to the interval where the instantaneous current exceeds 30% of the peak, used as the fit window. (c) Measured longitudinal phase space of a representative shot. Energy centroid (white), global linear fit (yellow dashed), and local fit within the high-current window (red) are used to extract energy chirp. (d) Shot-to-shot variation of the measured charge over the normal operation. (e, f) Distributions of extracted energy chirp (dE/dt) from global (e) and local (f) linear fitting, comparing quasi-flattop and standard laser profiles as the benchmark.*

Building on this, we now consider how spatiotemporal shaping could be extended beyond the flattop case toward more general beam control strategies. While this work focuses on flattop shaping at two representative charge regimes, the underlying method is inherently flexible in time and space dynamics. With appropriate laser modulation, the system could support a broader class of beam current profiles, such as ramps, double-humps, asymmetric shapes, or mode-lock behaviors, opening possibilities for tailored beam delivery in XFELs and other advanced beam-driven applications[10,44,52].

This may also offer compatibility with high-repetition-rate and superconducting linacs: a promising route toward real-time, feedback-enabled optimization[70–73]. As large-scale user facilities transition to MHz-rate operation, the ability to dynamically adapt the electron beam at the source, without extensive downstream retuning (compared with lower charge settings), may prove increasingly essential. Beyond FELs, the same approach could be extended to other photocathode‑based systems where deterministic phase‑space control remains a central challenge. For example, at SPring‑8, the use of 3D–shaped UV laser pulses with a flattop spatial profile combined with a flattop temporal profile of around 9 ps, reduced normalized emittance down to ~1.2 π·mm·mrad at high bunch charge, significantly enhancing the beam quality needed for UED‑level temporal resolution[74]. Similarly, in ICS, the laser pulse optimization via two oppositely chirped pulses with controlled delay and phase has been shown to suppress non‑linear spectral broadening in the backscattered X- or γ‑ray output, yielding a much narrower bandwidth which is critical for high‑resolution spectroscopy and imaging applications[75].

Our proposed method could be a setting for further exploration into more complex beam shaping scenarios, where customizable spatiotemporal laser fields could be used to modulate longitudinal phase space with a feedback loop and precision. This opens a pathway not only to performance enhancement, but to functional beam design at the source—potentially enabling new operating modes, automated optimization, novel pump–probe geometries, or control over beam-plasma interactions.

We have experimentally demonstrated quasi-flattop laser shaping at the LCLS-II, enabling direct control over the longitudinal structure of high-brightness electron beams using spatiotemporally shaped UV pulses. By systematically varying the shaping slope via simulations, we found that moderate asymmetry ($K \approx -0.1 \sim -0.3$) not only minimizes emittance growth but also compensates for photoinjector-induced distortion, yielding a flatter output current profile with sharper rise and fall edges. This shaping leads to reduced energy chirp and enhanced current uniformity across different charge levels. At low charge, chirp is suppressed, but stability is limited by weaker collective effects and sensitivity to profile variations. At high charge, where space-charge and compression dominate, the same shaping significantly improves both chirp and shot-to-shot consistency, even without a laser heater. These results demonstrate that quasi-flattop shaping can effectively compensate for injector dynamics and improve beam quality in realistic operating conditions. This approach utilizes spatiotemporal ultrafast laser shaping of the photocathode drive laser, allowing for seamless integration into existing photoinjectors without the need for complex lattice changes. It provides a precise upstream knob for linearizing phase space and optimizing current uniformity in XFELs and other advanced electron sources.

## Methods

### UV Laser shaping

We previously introduced a methodology referred to as dispersion-controlled nonlinear synthesis (DCNS)[52], which expands upon the underlying concept of spectral compression. This technique involves taking a transform-limited broadband optical pulse, such as one spanning several nanometers, and dividing it into two identical replicas. Each pulse is then imparted with a substantial amount of second-order dispersion (SOD), but in equal and opposite directions. The two dispersed pulses are subsequently combined using a non-collinear SFG setup, resulting in the production of a near-transform-limited pulse with picosecond duration. DCNS enhances this approach by introducing third-order dispersion (TOD) in addition to the SOD and incorporating spectral amplitude shaping before nonlinear interaction. Generating a narrowband SFG pulse with a desired shape requires careful control of phase terms. First, the input pulses must have equal and opposite second-order dispersion (SOD, $\varphi_2$), calculated using:

$$\Delta t = t_0 \sqrt{1 + \left(4 \ln 2 \frac{\varphi_2}{t_0^2}\right)^2} \tag{1}$$

where $t_0$ is the TL input duration, and $\Delta t$ is the target output duration. Third-order dispersion (TOD, $\varphi_3$) can be tuned more freely to shape the output: symmetric profiles result from equal and opposite $\varphi_3$, while asymmetry arises when the sign or magnitude differs.

The near transform-limited UV flattop pulse is generated from a 250-fs, 1024 nm Yb: KGW laser. The laser system is a 40 W Light Conversion Carbide. This system can maintain 40 W of average power at repetition rates between 100 kHz and 1 MHz, adjusting the pulse energy between 400 µJ and 40 µJ as required. The central wavelength is ~1024 nm with a bandwidth of 8 nm at 40 µJ and 7 nm at 400 µJ. The applied SOD and TOD were ±2.561 ps² and ∓0.28 ps³, respectively.

### Beamline Configuration

The LCLS-II soft X-ray (SXR) branch is based on a superconducting RF (SRF) linac capable of operating in continuous-wave (CW) mode at MHz repetition rates. The electron beam is generated in a normal-conducting very high frequency (VHF)-band photoinjector gun, where a UV laser pulse illuminates the photocathode to release electrons. The gun is followed by a buncher cavity and focusing solenoids that provide initial longitudinal compression and transverse matching of the beam. This injector system has been described in detail here[3,76].

Downstream of the injector, 35 superconducting cryomodules accelerate the beam to energies up to 4 GeV. Two magnetic chicanes (BC1 and BC2), placed after the L1 and L2 linac sections, provide bunch compression to achieve kiloampere peak currents. The beam is then transported to the SXR undulator line, which generates femtosecond X-ray pulses in the 0.25–5 keV photon energy range. More details can be found in the supplementary.

**Simulation Modeling**

The beam dynamics simulations were performed using the Lightsource unified modeling environment (LUME) framework, which integrates the 3D space-charge tracking code IMPACT-T for the injector section and Genesis for free-electron laser interactions. The modeled beamline faithfully reproduces the LCLS-II photoinjector lattice, consisting of a 187 MHz CW normal-conducting RF gun, followed by a 1.3 GHz two-cell buncher cavity, two focusing solenoids, and a matching section of quadrupole and corrector magnets. The electron beam is tracked from the photocathode surface through the first superconducting cryomodule (CM01), reaching a final energy of 100 MeV at a position approximately 15 meters downstream. To ensure numerical convergence and accurately capture collective effects, simulations utilized 300,000 macroparticles per bunch. A 3D space-charge algorithm based on a shifted Green function method was active throughout the entire acceleration, accounting for the complex interplay between internal space-charge fields and external RF focusing. Cylindrical symmetry was assumed for the initial particle distribution to optimize computational efficiency while maintaining physical accuracy for the baseline dynamics. Simulated output parameters—including emittance, bunch length, current profile, and longitudinal phase space—are evaluated at the injector exit. These results are used to quantify the effects of laser temporal asymmetry on beam quality and identify optimal shaping conditions.

**Synthesis of Quasi-flattop Laser Profiles**

Transversely, the laser profile was modeled as a radially uniform "top-hat" distribution with a hard-edge cutoff at $R_{max} = 0.5$ mm to decouple longitudinal effects from transverse coupling variations. Temporally, the study mapped the transition from theoretical idealism to experimental reality by synthesizing profiles ranging from a standard Gaussian and an ideal flattop to a series of "quasi-flattop" profiles. These asymmetric profiles were parameterized by a normalized slope ranging from -0.15 to -0.50. To satisfy charge conservation while maintaining a fixed FWHM of 16.5 ps, the pulse geometry was allowed to evolve self-consistently.

**Diagnostics and Data Acquisition**

The temporal profile of the UV pulse is characterized using third-harmonic cross-correlation with a 70 fs, 1035 nm pulse sourced from the oscillator that drives the main laser amplifier. To monitor X-ray pulse energy on a shot-by-shot basis, gas-monitor detectors (GMDs) utilize photoionization of rare gases at low pressure, measuring the resulting photoions and photoelectrons.

Diagnostics based on transverse deflecting cavities (TDCs) are employed to examine the longitudinal phase space of the electron beam. These are positioned both upstream along the superconducting beamline and downstream of the soft and hard X-ray undulators. S-band TDCs (STCAV) are situated after the laser heater, before the main acceleration in the linac, while X-band TDCs (XTCAV) are placed after each undulator line. All the data are collected by the customized data-acquisition GUI. More details are available in the supplementary information.

**TCAV Calibrations**

The TCAV data calibration process begins by calculating the temporal resolution per degree,

$$t_{deg} = \frac{\lambda_s}{360\,[deg] * c}\,[s/deg]$$

using the streaking cavity wavelength ($\lambda_s$) and the speed of light ($c$), which is then used to convert pixel measurements to femtoseconds via incorporating the screen resolution (res) and streaking amount (streak) with resolution data extracted from files (ProfMon-OTRS_DIAG0):

$$px2fs = \frac{t_{deg}\,[s/deg] * res\,[\mu m/pixel]}{streak\,[\mu m/deg]} * 1e15$$

Simultaneously, the energy axis is determined by

$$px2MeV = \frac{res\,[\mu m/pixel]}{dispersion\,[\mu m/MeV]} * Beam\ energy$$

**Extraction of energy chirp from phase space**

To extract the energy chirp, we first compute the energy centroid $\underline{E}(t)$ from the phase-space image $W(E, t)$. For each time slice, we locate the peak and compute the centroid within a narrow window (~4 pixels) around the peak to suppress the noise. This centroid is smoothed (window = 11) and interpolated over gaps. Weighted linear regression on $\underline{E}(t)$ has been applied to obtain $dE/dt$, using the beam current as weights. Two fits are performed:
- Global fitting: overall valid time points above the threshold;
- Local fitting: in a window with half-width (12 fs).

The longitudinal chirp is defined as:

$$h = \frac{1}{E_0}\frac{dE}{dz} \approx \frac{1}{cE_0}\frac{dE}{dt},$$

where $E_0$ is the energy and $c$ is the speed of light.

## Data availability

All other data that support the plots within this paper and other findings of this study are available from the corresponding authors on reasonable request.

## Acknowledgements

The authors would like to acknowledge the support from the SLAC National Accelerator Laboratory, the U.S. Department of Energy (DOE), the Office of Science, Office of Basic Energy Sciences under Contract No. DE-AC02-76SF00515, No. DE-SC0022559, No. DE-FOA-0002859, the National Science Foundation under Contract Nos. 2231334, 2431903, and 2436343, and AFOSR Contract No. FA9550-23-1-0409. The authors would like to thank the useful discussions with Zhirong Huang about data calibrations.

## Contributions

H.Z., R.L., J.H., set up the UV laser shaping experiment, photoinjector preparations and diagnostics. H.Z., R.L., J.H., N.N., R.R., P.F., D.S., N.S., K.B., Z.Z., K.L., C.P., R.O., V.G., Z.H., G.J., J.C., A.M., and S.C. conducted the XFEL beamline experiments and data collections. M.B., J.H., and H.Z., set up the FROG. H.Z., J.H., N.N., B.M., J.B., J.Q., S.C., performed the data analysis. All authors were involved in the writing of the manuscript.

## Competing interests

The authors declare no competing interests.

## Supplementary Information

- Photoinjector laser shaping system
- Diagnostic system
- Beamline configuration for different runs
- Charge region calculation
- Cavity stability measurements

## Reference

1. Yan, J. *et al.* Terawatt-attosecond hard X-ray free-electron laser at high repetition rate. *Nat. Photonics* **18**, 1293–1298 (2024).

2. McNeil, B. & Thompson, N. X-ray free-electron lasers. *Nature Photonics* **4**, 814–821 (2010).

3. Zhang, H. *et al.* The Linac Coherent Light Source II photoinjector laser infrastructure. *High Power Laser Science and Engineering* **12**, e51 (2024).

4. Nam, I. *et al.* High-brightness self-seeded X-ray free-electron laser covering the 3.5 keV to 14.6 keV range. *Nat. Photonics* **15**, 435–441 (2021).

5. Liu, S. *et al.* Cascaded hard X-ray self-seeded free-electron laser at megahertz repetition rate. *Nat. Photonics* **17**, 984–991 (2023).

6. Baum, P. & Zewail, A. H. Breaking resolution limits in ultrafast electron diffraction and microscopy. *Proc. Natl. Acad. Sci. U. S. A.* **103**, 16105–16110 (2006).

7. Yang, J. *et al.* Simultaneous observation of nuclear and electronic dynamics by ultrafast electron diffraction. *Science* **368**, 885–889 (2020).

8. Graves, W. S., Bessuille, J., Brown, P. & Carbajo, S. Compact x-ray source based on burst-mode inverse Compton scattering at 100 kHz. *Physical Review Special* (2014).

9. Gadjev, I. *et al.* An inverse free electron laser acceleration-driven Compton scattering X-ray source. *Sci. Rep.* **9**, 532 (2019).

10. Lindroth, E. *et al.* Challenges and opportunities in attosecond and XFEL science. *Nat. Rev. Phys.* **1**, 107–111 (2019).

11. Guo, Z. *et al.* Experimental demonstration of attosecond pump–probe spectroscopy with an X-ray free-electron laser. *Nat. Photonics* **18**, 691–697 (2024).

12. Franz, P. *et al.* Terawatt-scale attosecond X-ray pulses from a cascaded superradiant free-electron laser. *Nat. Photonics* **18**, 698–703 (2024).

13. Wan, Y. *et al.* Femtosecond electron microscopy of relativistic electron bunches. *Light Sci. Appl.* **12**, 116 (2023).

14. Zhang, H. *et al.* High-endurance micro-engineered LaB6 nanowire electron source for high-resolution electron microscopy. *Nat. Nanotechnol.* **17**, 21–26 (2022).

15. Wang, W. T. *et al.* High-brightness high-energy electron beams from a laser Wakefield accelerator via energy chirp control. *Phys. Rev. Lett.* **117**, 124801 (2016).

16. Winkler, P. *et al.* Active energy compression of a laser-plasma electron beam. *Nature* **640**, 907–910 (2025).

17. Yakimenko, V. *et al.* Prospect of studying nonperturbative QED with beam-beam collisions. *Phys. Rev. Lett.* **122**, 190404 (2019).

18. Krasilnikov, M. *et al.* First high peak and average power single-pass THz free-electron laser in operation. *Phys. Rev. Accel. Beams* **28**, (2025).

19. Gu, Y.-J., Klimo, O., Bulanov, S. V. & Weber, S. Brilliant gamma-ray beam and electron–positron pair production by enhanced attosecond pulses. *Commun. Phys.* **1**, (2018).

20. Qiang, J. Fast longitudinal beam dynamics optimization in x-ray free electron laser linear accelerators. *Phys. Rev. Accel. Beams* **22**, (2019).

21. Lindstrøm, C. A. & Thévenet, M. Emittance preservation in advanced accelerators. *J. Instrum.* **17**, P05016 (2022).

22. Lindstrøm, C. A. *et al.* Emittance preservation in a plasma-wakefield accelerator. *Nat. Commun.* **15**, 6097 (2024).

23. Emma, P., Huang, Z., Kim, K.-J. & Piot, P. Transverse-to-longitudinal emittance exchange to improve performance of high-gain free-electron lasers. *Phys. Rev. ST Accel. Beams* **9**, 100702 (2006).

24. Prat, E. *et al.* Emittance measurements and minimization at the SwissFEL Injector Test

Facility. *Phys. Rev. Spec. Top. - Accel. Beams* **17**, (2014).

25. Prat, E. *et al.* Generation and characterization of intense ultralow-emittance electron beams for compact X-ray free-electron lasers. *Phys. Rev. Lett.* **123**, 234801 (2019).

26. Qiang, J. *et al.* Start-to-end simulation of x-ray radiation of a next generation light source using the real number of electrons. *Phys. Rev. Spec. Top. - Accel. Beams* **17**, (2014).

27. Neveu, N. *et al.* Simulation of nonlinearly shaped UV pulses in LCLS-II. *Nucl. Instrum. Methods Phys. Res. A* **1072**, 170065 (2025).

28. Penco, G. *et al.* Experimental demonstration of electron longitudinal-phase-space linearization by shaping the photoinjector laser pulse. *Phys. Rev. Lett.* **112**, 044801 (2014).

29. Zhu, Z. *et al.* Inhibition of current-spike formation based on longitudinal phase space manipulation for high-repetition-rate X-ray FEL. *Nucl. Instrum. Methods Phys. Res. A* **1026**, 166172 (2022).

30. Huang, Z. & Kim, K.-J. Review of x-ray free-electron laser theory. *Phys. Rev. Spec. Top. - Accel. Beams* **10**, (2007).

31. Wu, Y. *et al.* Linearization of an electron beam's longitudinal phase space using a hollow-channel plasma. *Phys. Rev. Appl.* **19**, (2023).

32. Jia, B., Wu, Y. K., Bisognano, J. J., Chao, A. W. & Wu, J. Influence of an imperfect energy profile on a seeded free electron laser performance. *Phys. Rev. Spec. Top. - Accel. Beams* **13**, (2010).

33. Bettoni, S. *et al.* Impact of laser stacking and photocathode materials on microbunching stability in photoinjectors. *Phys. Rev. Accel. Beams* **23**, (2020).

34. Kretschmar, M. *et al.* Compact realization of all-attosecond pump-probe spectroscopy. *Sci.*

*Adv.* **10**, eadk9605 (2024).

35. Midorikawa, K. Progress on table-top isolated attosecond light sources. *Nat. Photonics* **16**, 267–278 (2022).

36. Li, S. *et al.* Attosecond-pump attosecond-probe x-ray spectroscopy of liquid water. *Science* **383**, 1118–1122 (2024).

37. Di Mitri, S., Cornacchia, M., Spampinati, S. & Milton, S. Suppression of microbunching instability with magnetic bunch length compression in a linac-based free electron laser. *Phys. Rev. Spec. Top. - Accel. Beams* **13**, (2010).

38. Musumeci, P., Li, R. K., Roberts, K. G. & Chiadroni, E. Controlling nonlinear longitudinal space charge oscillations for high peak current bunch train generation. *Phys. Rev. Spec. Top. - Accel. Beams* **16**, (2013).

39. Zhang, Z., Ding, Y., Huang, Z. & Zhou, F. Multiplexed photoinjector optimization for high-repetition-rate free-electron lasers. *Frontiers in Physics* **11**, (2023).

40. Götzfried, J. *et al.* Physics of high-charge electron beams in laser-plasma wakefields. *Phys. Rev. X.* **10**, (2020).

41. Rosenzweig, J. B. *et al.* Ultra-high brightness electron beams from very-high field cryogenic radiofrequency photocathode sources. *Nucl. Instrum. Methods Phys. Res. A* **909**, 224–228 (2018).

42. Prat, E. *et al.* Experimental demonstration of mode-coupled and high-brightness self-amplified spontaneous emission in an X-ray free-electron laser. *Phys. Rev. Lett.* **133**, 205001 (2024).

43. Galletti, M. *et al.* Prospects for free-electron lasers powered by plasma-wakefield-

accelerated beams. *Nat. Photonics* **18**, 780–791 (2024).

44. Hu, W. *et al.* Tunable, phase-locked hard X-ray pulse sequences generated by a free-electron laser. *arXiv [physics.acc-ph]* (2025).

45. Duris, J. *et al.* Tunable isolated attosecond X-ray pulses with gigawatt peak power from a free-electron laser. *Nat. Photonics* **14**, 30–36 (2019).

46. Cho, D. H. *et al.* High-throughput 3D ensemble characterization of individual core-shell nanoparticles with X-ray free electron laser single-particle imaging. *ACS Nano* **15**, 4066–4076 (2021).

47. Wang, G., Dijkstal, P., Reiche, S., Schnorr, K. & Prat, E. Millijoule femtosecond X-ray pulses from an efficient fresh-slice multistage free-electron laser. *Phys. Rev. Lett.* **132**, 035002 (2024).

48. Baumgärtner, K. *et al.* Ultrafast orbital tomography of a pentacene film using time-resolved momentum microscopy at a FEL. *Nat. Commun.* **13**, 2741 (2022).

49. Chen, Z. *et al.* Ultrafast multi-cycle terahertz measurements of the electrical conductivity in strongly excited solids. *Nat. Commun.* **12**, 1638 (2021).

50. Hemsing, E., Stupakov, G., Xiang, D. & Zholents, A. Beam by design: Laser manipulation of electrons in modern accelerators. *Rev. Mod. Phys.* **86**, 897–941 (2014).

51. Kling, M. F. *et al.* Roadmap on basic research needs for laser technology. *J. Opt.* (2024) doi:10.1088/2040-8986/ad8458.

52. Lemons, R., Neveu, N., Duris, J., Marinelli, A. & Durfee, C. Temporal shaping of narrow-band picosecond pulses via noncolinear sum-frequency mixing of dispersion-controlled pulses. *Review Accelerators and …* (2022).

53. Nielsen, C. E. & Sessler, A. Longitudinal space charge effects in particle accelerators. *Review of Scientific Instruments* **30**, 80–89 (1959).

54. Fisher, A. *et al.* Single-pass high-efficiency terahertz free-electron laser. *Nat. Photonics* **16**, 441–447 (2022).

55. Antipov, S. A., Gubaidulin, V., Agapov, I., Cortés García, E. C. & Gamelin, A. Space charge effects in fourth-generation light sources: The PETRA IV and SOLEIL II cases. *Phys. Rev. Accel. Beams* **28**, (2025).

56. Musumeci, P., Moody, J. T., England, R. J., Rosenzweig, J. B. & Tran, T. Experimental generation and characterization of uniformly filled ellipsoidal electron-beam distributions. *Phys. Rev. Lett.* **100**, 244801 (2008).

57. Hirschman, J. *et al.* Leveraging the capabilities of LCLS-II: linking adaptable photoinjector laser shaping to x-ray diagnostics through start-to-end simulation. *meow.elettra.eu* Preprint at https://doi.org/10.18429/JACOW-IPAC2025-MOPB040 (2025).

58. Ilia, D. *et al.* Novel photoinjector laser providing advanced pulse shaping for FLASH and EuXFEL. *meow.elettra.eu* Preprint at https://doi.org/10.18429/JACOW-IPAC2025-THPB027 (2025).

59. Sharma, A. K., Tsang, T. & Rao, T. Theoretical and experimental study of passive spatiotemporal shaping of picosecond laser pulses. *Phys. Rev. Spec. Top. - Accel. Beams* **12**, (2009).

60. Lemons, R., Hirschman, J., Zhang, H., Durfee, C. & Carbajo, S. Nonlinear Shaping in the Picosecond Gap. *Ultrafast Science* **4**, 0112 (2025).

61. Cialdi, S., Vicario, C., Petrarca, M. & Musumeci, P. Simple scheme for ultraviolet time-pulse

shaping. *Appl. Opt.* **46**, 4959–4962 (2007).

62. Weiner, A. Femtosecond pulse shaping using spatial light modulators. *Rev. Sci. Instrum.* **71**, 1929–1960 (2000).

63. Will, I. Generation of flat-top picosecond pulses by means of a two-stage birefringent filter. *Nucl. Instrum. Methods Phys. Res. A* **594**, 119–125 (2008).

64. Weiner, A. M. Ultrafast optical pulse shaping: A tutorial review. *Opt. Commun.* **284**, 3669–3692 (2011).

65. Frantz, L. M. & Nodvik, J. S. Theory of pulse propagation in a laser amplifier. *J. Appl. Phys.* **34**, 2346–2349 (1963).

66. Hirschman, J., Lemons, R., Wang, M., Kroetz, P. & Carbajo, S. Design, tuning, and blackbox optimization of laser systems. *Opt. Express* (2022) doi:10.1364/oe.520542.

67. Lemons, R., Hirschman, J., Zhang, H., Durfee, C. & Carbajo, S. Nonlinear shaping in the picosecond gap. *Ultrafast Sci* (2025) doi:10.34133/ultrafastscience.0112.

68. Emma, P. *et al.* First lasing and operation of an ångstrom-wavelength free-electron laser. *Nat. Photonics* **4**, 641–647 (2010).

69. Li, S. *et al.* 'Beam `a la carte': laser heater shaping for attosecond pulses in a multiplexed x-ray free-electron laser. *Applied Physics Letters* **125**, (2024).

70. Mishra, A. *et al.* A start to end Machine Learning approach to maximize scientific throughput from the LCLS-II-HE. *arXiv [physics.ins-det]* (2025).

71. Ji, F. *et al.* Multi-objective Bayesian active learning for MeV-ultrafast electron diffraction. *Nat. Commun.* **15**, 4726 (2024).

72. Roussel, R. *et al.* Turn-key constrained parameter space exploration for particle

accelerators using Bayesian active learning. *Nat. Commun.* **12**, 5612 (2021).

73. Li, K. *et al.* Prediction on X-ray output of free electron laser based on artificial neural networks. *Nat. Commun.* **14**, 7183 (2023).

74. Tomizawa, H. *et al.* Adaptive 3-d UV-laser pulse shaping system to minimize emittance for photocathode Rf gun and new laser incidence system. *Proceedings of FEL* (2007).

75. Seipt, D., Kharin, V. Y. & Rykovanov, S. G. Optimizing laser pulses for narrow-band inverse Compton sources in the high-intensity regime. *Phys. Rev. Lett.* **122**, 204802 (2019).

76. Zhou, F. *et al.* Commissioning of the SLAC Linac Coherent Light Source II electron source. *Phys. Rev. Accel. Beams* **24**, (2021).

**Supplementary note 1: Photoinjector laser shaping system**

As we introduced by our previous works [1-2], the IR front-end employs a Ytterbium-based chirped pulse amplification (CPA) system. It delivers pulse energies up to 50 µJ at a central wavelength of 1030 nm, operating at repetition rates as high as 1 MHz, with a transform-limited pulse width of about 330 fs [3]. The output pulse duration can be tuned continuously over a wide range, from ~330 fs up to ~20 ps. The CPA chain is seeded by a mode-locked oscillator that is phase-locked to the facility RF reference, ensuring synchronization of the laser system with the accelerator timing. For temporal diagnostics, such as measurements with an optical cross-correlator, a short-pulse branch is implemented. To enable advanced control over the temporal profile, the laser incorporates a programmable spectral phase and amplitude shaper, which allows tailoring of the spectrum and customization of shaping of the UV pulses.

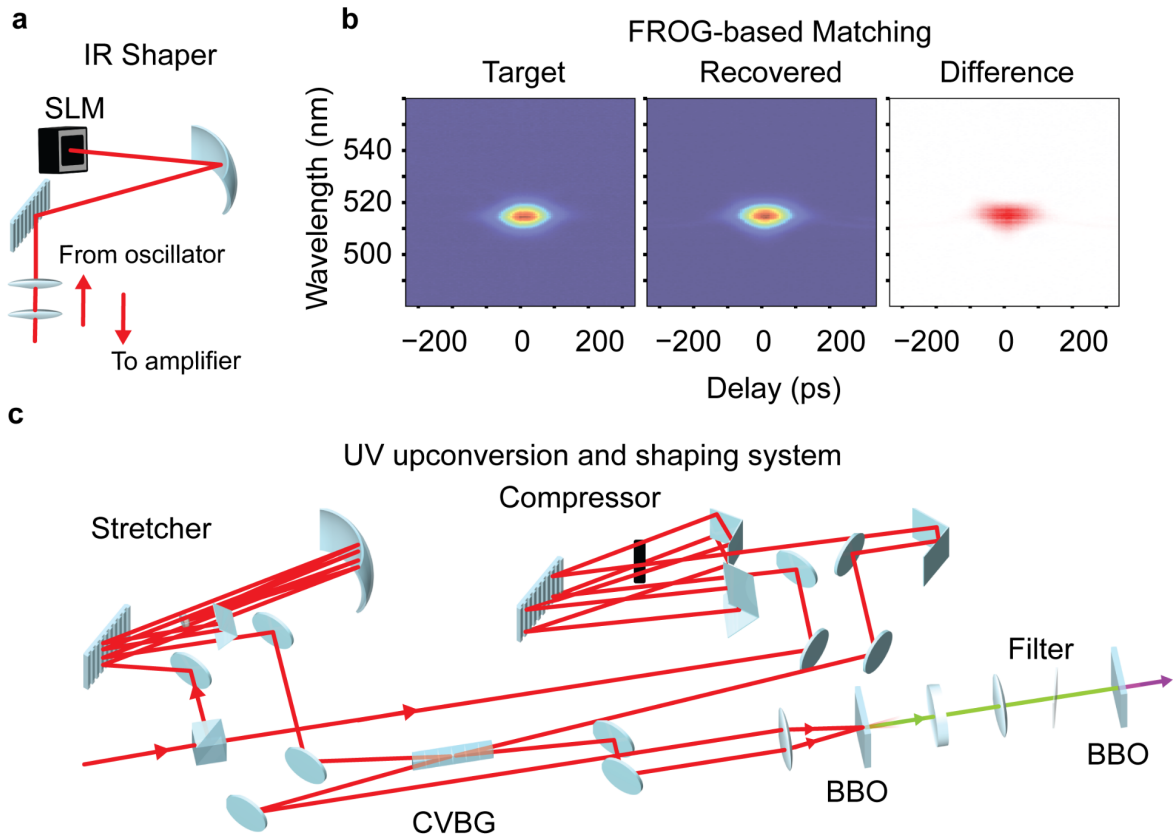

Figure S1. (a) Photoinjector laser IR shaping setup. (b) FROG tuning illustrations and (c) DCNS upconversion and passive laser shaping setup.

To enable flexible temporal shaping, the system includes a programmable spectral phase and amplitude modulator. For photoinjector applications, a spatial light modulator (SLM, Santec SLM-210 reflective liquid-crystal modulator, 1920 × 1200 pixels, 15-mm active length) in a grating-based (1379 grooves/mm) folded 4f arrangement (Figure S1a) is used to apply a phase mask at the Fourier plane. The shaped pulses are then directed into a regenerative chirped-pulse amplifier. This commercial system can deliver up to 40 W average power at 1025 nm, producing 256 fs pulses at repetition rates up to 1 MHz under standard operating conditions. A custom-built frequency-resolved optical gating (FROG) system was employed to record spectrograms of the shaped IR pulses before amplification. These measurements reveal both the spectral content and temporal structure of the pulses. The spectrograms were used to identify appropriate SLM configurations that pre-compensate residual dispersion unintentionally introduced by the 4f geometry of the shaping setup (see Figure S1b).

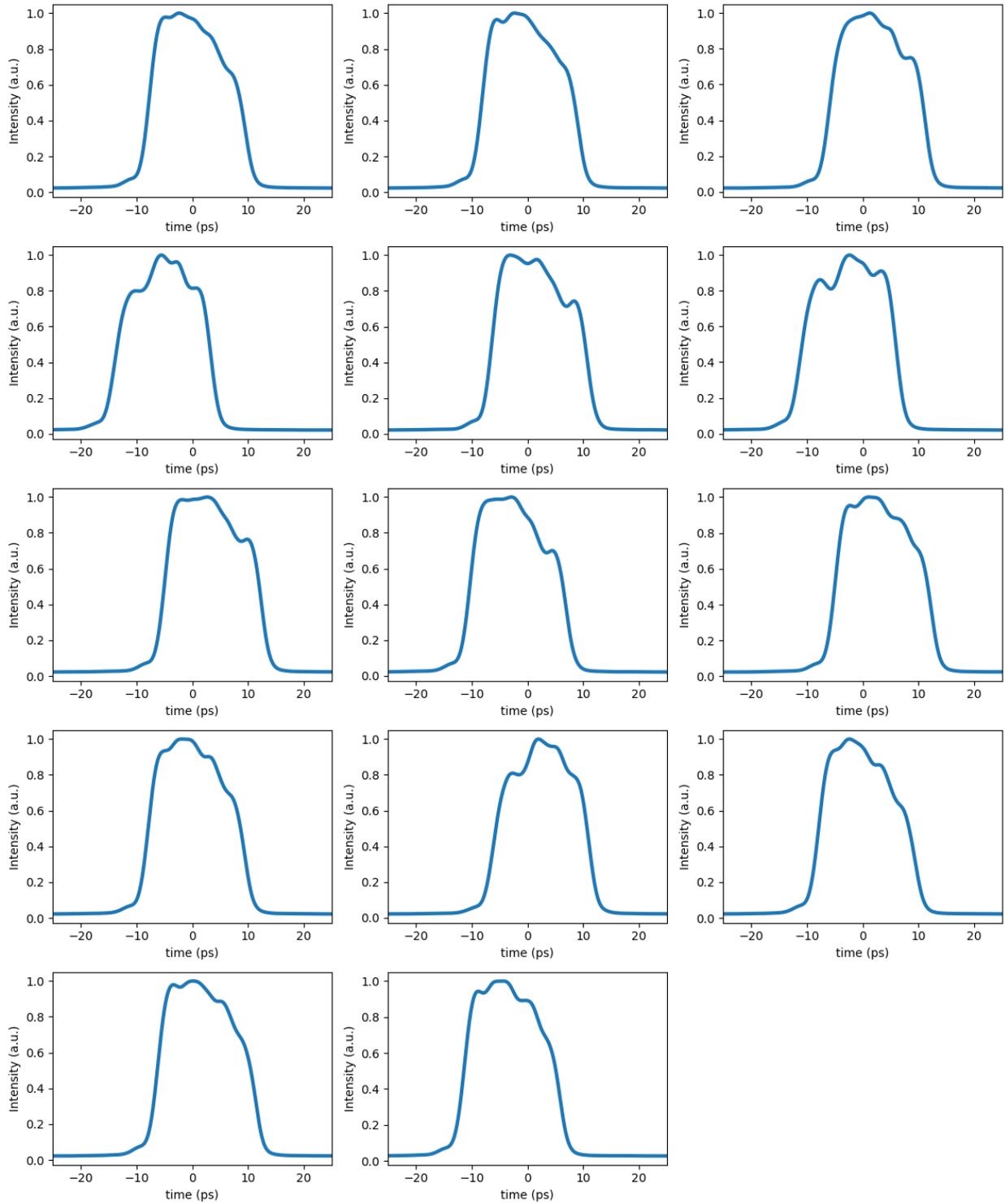

Figure S2. Quasi-flattop UV temporal profile with different SLM settings.

The amplified IR pulses are subsequently converted to the UV. Several nonlinear upconversion approaches exist, including SHG and SFG. The amplified pulse is split into two replicas, with equal and opposite second- and third-order dispersion applied using a combination of compressor and stretcher optics together with a chirped volume Bragg grating (CVBG) (Figure S1c). The two pulses are then recombined via noncollinear SFG in a BBO crystal to produce green light, which is further doubled to the UV in a second BBO stage. In the experimental setup, additional amplitude masks are placed near the Fourier plane of the compressor line to suppress unwanted spectral components and improve the output quality. The generated UV beam is guided from the laser room (sector 0) to the LCLS-II injector tunnel using evacuated beamlines, since the RF gun is located within a radiation-shielded enclosure. A spatially flattop laser profile was achieved on the cathode by relay imaging an iris placed in the beam path, which blocked the Gaussian tails and transmitted only the central plateau (as shown in Figure S3).

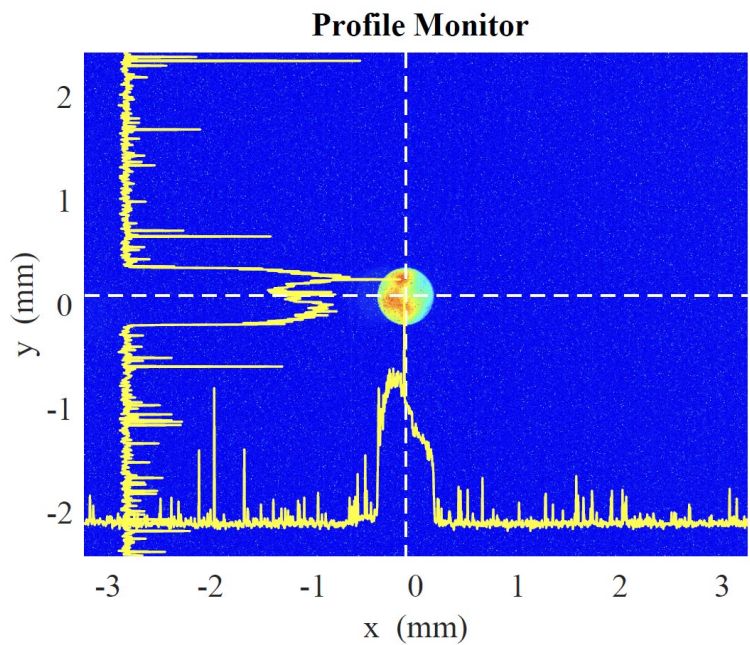

Figure S3. Laser profile before injecting into the cathode.

**Supplementary note 2: Diagnostic systems**

A downstream Ce: YAG (cerium-doped yttrium aluminium garnet) crystal screen is recorded with a camera to record the transverse profile of the electron beam. The horizontal direction of the image corresponds to time, while the vertical direction reflects beam energy. Time calibration is performed by tracking how the beam centroid shifts with the RF phase of the X-band deflector, and the vertical scale is referenced against known changes in beam energy. A temporal resolution on the order of 1–3 fs (r.m.s.) can be achieved, depending on the beam energy.

At LCLS-II, the current S-band (2.856 GHz) RF transverse deflector yields about 10 fs r.m.s. resolution at a 5 GeV beam energy. The resolution depends on factors such as the operating frequency, deflecting voltage, and intrinsic beam properties. To push toward finer resolution, an X-band (11.424 GHz) transverse deflector was installed and later developed with dedicated structures and couplers. Because the X-band frequency is four times higher than that of the S-band system, it can sustain stronger RF fields, enabling a faster and more powerful sweep of the beam, thereby improving temporal resolution for the compressed beams.

The transverse emittance of the electron beam was determined using the standard quadrupole scan method. In this procedure, the focusing strength of a quadrupole magnet located upstream of a scintillator or optical transition radiation (OTR) screen is varied, and the resulting beam spot size at the screen is measured for each setting. The beam size squared, $\varepsilon_x^2$, depends quadratically on the quadrupole strength, and by fitting these data with the transport matrix formalism, the Twiss parameters $(\alpha, \beta, \gamma)$ and the projected emittance $\varepsilon$ are extracted. From this, the normalized emittance is obtained as $\varepsilon_n = \beta_{rel} \gamma_{rel} \varepsilon$.

**Supplementary note 3: Beamline configuration for different runs**

Table S1. The relevant machine parameters of different runs.

|  | Gaussian (low charge level) | Flattop (low charge level) | Flattop (high charge level) |
|---|---|---|---|
| SOLN:GUNB:212:BACT (KG-m) | 0.0440 | 0.0445 | 0.0441 |
| SOLN:GUNB:823:BACT (KG-m) | 0.0246 | 0.0246 | 0.0242 |
| BEND:BC1B:200 (KG-m) | 0.793 | 0.793 | 0.793 |
| BEND:BC1B:400 (KG-m) | 0.790 | 0.790 | 0.790 |
| BEND:BC1B:600 (KG-m) | 0.790 | 0.790 | 0.790 |
| BEND:BC1B:800 (KG-m) | 0.790 | 0.790 | 0.790 |
| BEND:BC2B:200 (KG-m) | 2.421 | 2.421 | 2.421 |
| BEND:BC2B:500 (KG-m) | 2.421 | 2.421 | 2.421 |
| BEND:BC2B:600 (KG-m) | 2.421 | 2.421 | 2.421 |
| BEND:BC2B:900 (KG-m) | 2.421 | 2.421 | 2.421 |
| GUN:GUNB:100:AACT (MV) | 0.751 | 0.751 | 0.751 |
| GUN:GUNB:100:PACT | 7.0 | 7.0 | 7.3 |
| ACCL:GUNB:455:PACT - BUNCH (deg) | 170.250 | 170.250 | 176.170 |
| ACCL:GUNB:455:AACT - amplitude (MV) | 0.134 | 0.134 | 0.134 |

## Supplementary note 4: Charge region calculation

The grey-shaded region in the beam current profile is quantitatively defined as the portion where the current density remains above 30% of the peak value, which highlights the central part of the distribution carrying the majority of the charge. Let $J(x)$ represent the measured current density at position x along the transverse direction of the beam. The maximum current density is defined as $J_{max}$. The threshold for defining the central region is then set as:

$$J_{th} = 0.3 * J_{max} \tag{1}$$

Only the positions where $J(x) \geq J_{th}$ are included in the shaded region. The boundaries of the region, denoted as $x_1$ and $x_2$, are obtained by solving the following equations:

$$J(x_1) = J(x_2) = J_{th} \tag{2}$$

where $x_1 <$ x_peak $< x_2$, and x_peak is the location of $J_{max}$. Thus, the central region $\Omega$ is formally defined as:

$$\Omega = \{x \mid J(x) \geq J_{th}\} \tag{3}$$

To verify that this criterion effectively captures the majority of the total beam charge, the normalized charge fraction η within the region is computed as:

$$\eta = \int_{x_1}^{x_2} J(x)dx / \int_{-\infty}^{+\infty} J(x)dx \tag{4}$$

For all profiles analyzed in this work, η consistently exceeded 70%, confirming that the selected region contains the dominant charge contribution.

**Supplementary note 5: Cavity stability measurements**

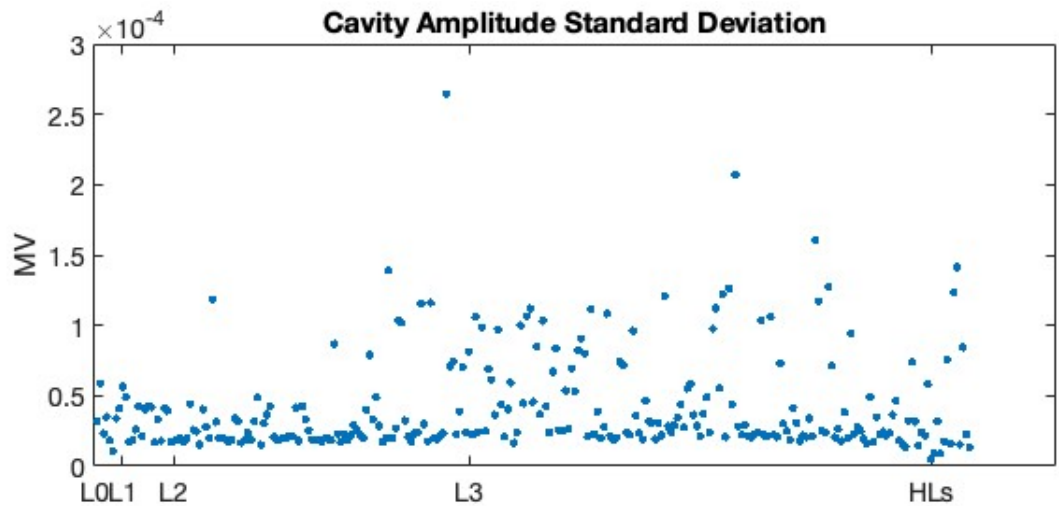

Figure S6. The amplitude stability measurements of different cavities.

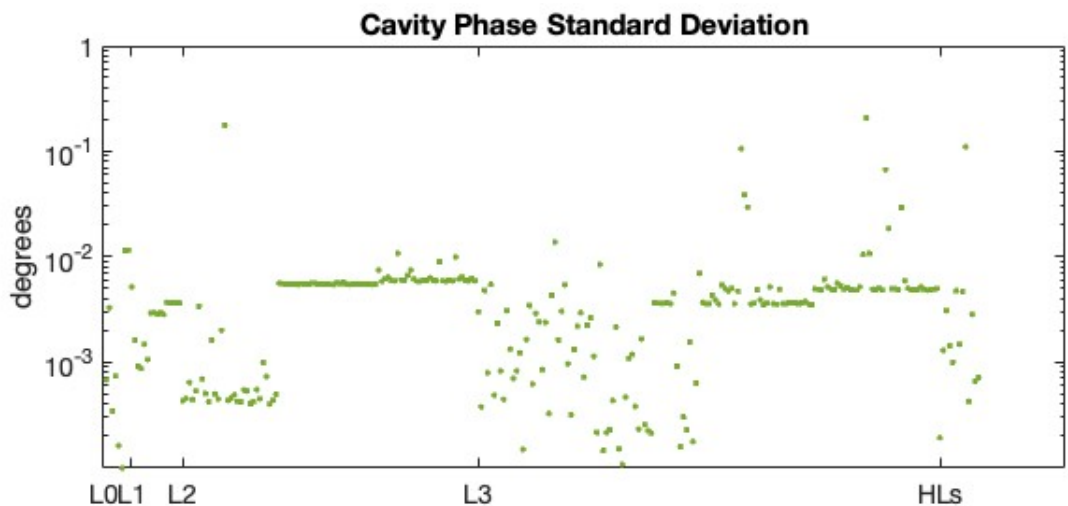

Figure S7. The phase stability measurements of different cavities.

[1] Lemons, Randy, et al. "Nonlinear shaping in the picosecond gap." Ultrafast Science 4 (2025): 0112.
[2] Lemons, Randy, et al. "Temporal shaping of narrow-band picosecond pulses via noncolinear sum-frequency mixing of dispersion-controlled pulses." Physical Review Accelerators and Beams 25.1 (2022): 013401.
[3] Zhang, Hao, et al. "The Linac coherent light source II photoinjector laser infrastructure." High Power Laser Science and Engineering 12 (2024): e51.